\documentclass[pre,aps,showpacs]{revtex4}
\usepackage{epsf}
\usepackage{amsmath}

\def\nn{\nonumber}
\def\beq{\begin{equation}}
\def\eeq{\end{equation}}
\def\bea{\begin{eqnarray}}
\def\eea{\end{eqnarray}}
\def\bi{\begin{itemize}}
\def\ei{\end{itemize}}
\def\be{\begin{enumerate}}
\def\ee{\end{enumerate}}

\begin{document}

\title{Relations between Potts and RSOS models on a torus}
\author{Jean-Fran\c{c}ois Richard${}^{1,2}$}
\author{Jesper Lykke Jacobsen${}^{1,3}$}
\affiliation{${}^1$LPTMS, Universit\'e Paris-Sud,
          B\^atiment 100, 91405 Orsay, France}
\affiliation{${}^2$LPTHE, Universit\'e Paris VI, Tour 16,
          4 place Jussieu, 75252 Paris cedex 05, France}
\affiliation{${}^3$Service de Physique Th\'eorique,
          CEA Saclay, 91191 Gif sur Yvette, France}
\date{\today}

\begin{abstract}

We study the relationship between $Q$-state Potts models and staggered RSOS models of
the $A_{p-1}$ type on a torus, with $Q^{1/2}=2\cos(\pi/p)$. In general the partition functions of these
models differ due to clusters of non-trivial topology. However we find exact
identities, valid for any temperature and any finite size of
the torus, between various modified partition functions in the two
pictures. The field theoretic interpretation of these modified partition
function is discussed.

\end{abstract}


\maketitle

\section{Introduction}
\label{sec1}

The two-dimensional Potts model \cite{Wu82,Saleur91} can be defined in terms of
integer valued spins $S_i=1,2,\ldots,Q$ living on the vertices $\{i\}$
of a lattice. Its partition function reads
\beq
 Z_{\rm spin} = \sum_{\{S_i\}} \prod_{\langle ij \rangle}
 {\rm e}^{K \delta(S_i,S_j)},
\eeq
where $\delta$ is the Kronecker delta and $\langle ij \rangle$ are the lattice
edges. In this paper we take the lattice to be an $L \times N$ square lattice
(say, of vertical width $L$ and horizontal length $N$) with toroidal boundary
conditions (i.e., periodic boundary conditions in both lattice directions).
We denote $V=L N$ the number of vertices of the lattice and $E=2 L N$
the number of edges.

This initial definition can be extended to arbitrary real values of $Q$
by means of a cluster expansion \cite{Kasteleyn}. One finds
\beq
 Z_{\rm cluster} = \sum_{\cal C} v^{e({\cal C})} Q^{c({\cal C})},
 \label{ZPotts}
\eeq
where $v={\rm e}^K-1$. Here, the sum is over the $2^E$
possible colourings ${\cal C}$ of the lattice edges (each
edge being either coloured or uncoloured), $e({\cal C})$ is the number
of coloured edges, and $c({\cal C})$ is the number of connected
components (clusters) formed by the coloured edges. For $Q$ a positive
integer one has $Z_{\rm spin} = Z_{\rm cluster}$.

Yet another formulation is possible when $Q^{1/2}=q+q^{-1}$ and
$q$ is a root of unity
\beq
 q = {\rm e}^{i \pi / p}, \qquad p=3,4,5,\ldots,
 \label{rootunity}
\eeq
this time in terms of a restricted height model with face interactions
\cite{ABF,Pasquier,PS}, henceforth referred to as the RSOS model. This
formulation is most easily described in an algebraic way. The Potts model
transfer matrix $T$ that adds one column of the square lattice can be written
in terms of the generators $e_j$ of the Temperley-Lieb algebra \cite{TL} as
follows
\bea
 T   &=& Q^{L/2} H_L \cdots H_2 H_1 V_L \cdots V_2 V_1,
 \label{TM} \\
 H_i &=& x I_{2i-1} + e_{2i-1}, \nn \\
 V_i &=& I_{2i} + x e_{2i}. \nn
\eea
Here, $H_i$ and $V_i$ are operators adding respectively horizontal and vertical
edges to the lattice, $I_j$ is the identity operator acting at site $j$,
and the parameter $x=Q^{-1/2} v = Q^{-1/2}({\rm e}^K-1)$.
The generators satisfy the well-known algebraic relations
\bea
 e_i e_{i\pm1} e_i &=& e_i, \nn \\
 (e_i)^2 &=& Q^{1/2} e_i, \\
 e_i e_j &=& e_j e_i \mbox{ for $|i-j| \ge 2$}.
\eea

\begin{figure}
\begin{center}
 \leavevmode
 \epsfysize=30mm{\epsffile{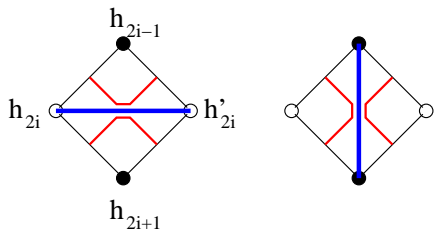}}
 \end{center}
 \protect\caption[3]{Graphical rendering of $V_i=I_{2i}+x e_{2i}$. Direct
(resp.\ dual) vertices are shown as full (resp.\ empty) circles. Coloured
edges (direct and dual) in the cluster picture are depicted as thick blue
lines. Their surrounding cluster boundaries are given as thin red lines.
RSOS heights $h_j$ are defined on both direct and dual vertices as shown.
The action of $I_{2i}$ (resp.\ $e_{2i}$) is illustrated on the left
(resp.\ right) part of the figure. Our convention is that the transfer
matrix acts towards the right.}
 \label{fig1}
\end{figure}

More precisely, $V_i$ can be thought of as adding a face to the lattice,
surrounded by two direct and two dual vertices, as shown in Fig.~\ref{fig1}.
$H_i$ is similarly defined, by exchanging direct and dual sites on the figure.

Meanwhile, the operators $I_j$ and $e_j$ can be represented in various ways,
thus giving rise to different transfer matrices.
When $Q$ is a positive integer, a spin representation of dimension $Q^L$
can be defined in an obvious way, and one has
\beq
 Z_{\rm spin} = {\rm Tr} \ (T_{\rm spin})^N.
\eeq

For any real $Q$ a cluster representation \cite{Blote82} of dimension
$C_L = \frac{(2L)!}{(L+1)! \, L!}$
can be defined by letting $I_j$ and $e_j$ act on the boundaries
\cite{BKW,BaxterBook} that separate direct and dual clusters, represented as
thin red lines in Fig.~\ref{fig1}. But it is impossible to write $Z_{\rm
cluster}$ as a trace of $T_{\rm cluster}$ defined in this
way. This is due to the existence of loops of cluster boundaries that are
non contractible with respect to the periodic boundary conditions in the
horizontal direction of the lattice.%
\footnote{It is however possible to modify the cluster representation
itself, and hence the transfer matrix, in a non-local way that allows to write
$Z_{\rm cluster}$ as a suitably modified trace \cite{Salas04}.}
For this reason, we shall not discuss $T_{\rm cluster}$ much further in
this paper, but we maintain Eq.~(\ref{ZPotts}) as the
definition of $Z_{\rm cluster}$ for real $Q$.

Finally, when $Q$ is given by Eq.~(\ref{rootunity}) the RSOS model is
introduced by letting $I_j$ and $e_j$ act on heights $h_j=1,2,\ldots,p-1$ defined
on direct and dual vertices \cite{ABF,Pasquier}. A pair of neighbouring
direct and dual heights are constrained to differ by $\pm 1$. In this
representation we have
\beq
 I_j = \delta(h_j,h'_j), \qquad
 e_j = \delta(h_{j-1},h_{j+1})
 \left(S_{h_j} S_{h'_j}\right)^{1/2} \left(S_{h_{j-1}}\right)^{-1},
 \label{weights}
\eeq
where $S_h = \sin(\pi h/p)$. Note that the clusters (direct and dual) are
still meaningful as they are surfaces of constant height. The constants $S_h$
are actually the components of the Perron-Frobenius eigenvector of the
incidence matrix (of size $p-1$)
\beq
 G_{p-1} = \left(
 \begin{tabular}{ccccc}
 0        & 1        & 0        & $\cdots$ & 0        \vspace{-0.2cm} \\
 1        & 0        & 1        &          & $\vdots$ \vspace{-0.2cm} \\
 0        & $\ddots$ & $\ddots$ & $\ddots$ & 0        \vspace{-0.2cm} \\
 $\vdots$ &          & 1        & 0        & 1        \vspace{0.05cm} \\
 0        & $\cdots$ & 0        & 1        & 0        \\
 \end{tabular} \right)
\eeq
of the Dynkin diagram $A_{p-1}$ \cite{Pasquier}. In
this representation, of dimension ${\rm Tr} \, (G_{p-1})^{2L} \sim Q^L$, we now
define the transfer matrix $T_{\rm RSOS}$ by Eq.~(\ref{TM}) and the partition
function $Z_{\rm RSOS} = {\rm Tr} \, (T_{\rm RSOS})^N$, the trace being over
allowed height configurations.

In this paper we discuss the relations between $Z_{\rm RSOS}$ and
$Z_{\rm cluster}$, with $Q^{1/2}=2 \cos(\pi/p)$ cf.~Eq.~(\ref{rootunity}).
These partition functions are in general different, due to clusters
of non-trivial topology wrapping around the torus.

We start by showing numerically, in section \ref{sec2}, that for $Q=3$ the
transfer matrices $T_{\rm spin}$ and $T_{\rm RSOS}$ have nonetheless many
identical eigenvalues. Defining various sectors (motivated by duality and
parity considerations) and also twisting the periodic boundary conditions in
different ways, we are able to conjecture several relations between the
corresponding transfer matrix spectra.

With this numerical motivation we then go on, in section~\ref{sec3},
to prove these relations on the level of the RSOS and cluster model
partition functions on finite
$L \times N$ tori. Some of the relations are specific to $Q=2$ ($p=4$) and
$Q=3$ ($p=6$), and some hold for general values of $p$. All of them hold
for arbitrary values of the temperature variable $x$. We stress that in
all cases the proofs are based on rigorous combinatorial considerations.

We conclude the paper, in section~\ref{sec4}, by interpreting our results,
and the various partition functions introduced, on the level of conformal
field theory, at the selfdual temperature $x=1$ where the Potts model is at
a critical point.

\section{Transfer matrix spectra}
\label{sec2}

The spectra of the transfer matrices $T_{\rm spin}$ and $T_{\rm RSOS}$ 
are easily studied numerically by exact diagonalisation techniques.
Denoting the eigenvalues as $\Lambda_i$, with $i=1,2,\ldots,{\rm dim}(T)$,
the results are most conveniently stated in terms of the corresponding
free energies per spin, $f_i=-L^{-1}\log(\Lambda_i)$.
Sample results for $Q=3$, width $L=2$, and temperature variable $x=5$ are shown
in Table~\ref{tab1}. 

\begin{table}
 \begin{tabular}{l|l|l|l|l}
 Transfer matrix: \ & $T_{\rm RSOS}^{\rm even}$ \ & $T_{\rm RSOS}^{\rm odd}$ \  &  $T_{\rm spin}$ & $T_{\rm spin}^{\rm dual}$ \\
 Twist: & $I$ \ $Z_2$ & $I$ \ $Z_2$ & $I$ \ $Z_2$ \ $Z_3$ \ & $I$ \ $Z_2$ \ $Z_3$ \\ \hline
-4.547135105405  & 1 \ \ 0 &  1 \ \ 0  & 1 \ \, 0 \ \, 0  & 1 \ \, 0 \ \, 0 \\
-4.536300662409  & 1 \ \ 0 &  1 \ \ 0  & 0 \ \, 1 \ \, 0  & 0 \ \, 1 \ \, 0 \\
-4.530748290953  & 1 \ \ 1 &  0 \ \ 0  & 2 \ \, 0 \ \, 0  & 0 \ \, 0 \ \, 1 \\
-3.512711596812  & 0 \ \ 0 &  1 \ \ 1  & 0 \ \, 0 \ \, 1  & 2 \ \, 0 \ \, 0 \\
-3.502223380184  & 1 \ \ 0 &  1 \ \ 0  & 0 \ \, 1 \ \, 0  & 0 \ \, 1 \ \, 0 \\
-3.441474985184  & 0 \ \ 0 &  0 \ \ 0  & 0 \ \, 0 \ \, 2  & 0 \ \, 0 \ \, 2 \\
-3.397645107750  & 0 \ \ 2 &  0 \ \ 2  & 0 \ \, 2 \ \, 0  & 0 \ \, 2 \ \, 0 \\
-3.348639214318  & 0 \ \ 0 &  0 \ \ 0  & 0 \ \, 0 \ \, 2  & 0 \ \, 0 \ \, 2 \\
-3.292754029664  & 1 \ \ 0 &  2 \ \ 1  & 0 \ \, 1 \ \, 1  & 2 \ \, 1 \ \, 0 \\
-2.335814864962  & 1 \ \ 0 &  1 \ \ 0  & 1 \ \, 0 \ \, 0  & 1 \ \, 0 \ \, 0 \\
-2.307465012288  & 2 \ \ 1 &  1 \ \ 0  & 2 \ \, 1 \ \, 0  & 0 \ \, 1 \ \, 1 \\
-2.285900912958  & 1 \ \ 0 &  1 \ \ 0  & 0 \ \, 1 \ \, 0  & 0 \ \, 1 \ \, 0 \\
-2.251579827634  & 0 \ \ 0 &  0 \ \ 0  & 0 \ \, 0 \ \, 2  & 0 \ \, 0 \ \, 2 \\
-2.236228400659  & 0 \ \ 0 &  1 \ \ 1  & 0 \ \, 0 \ \, 1  & 2 \ \, 0 \ \, 0 \\
-2.203480723895  & 1 \ \ 1 &  0 \ \ 0  & 2 \ \, 0 \ \, 0  & 0 \ \, 0 \ \, 1 \\
-2.202573934202  & 0 \ \ 2 &  0 \ \ 2  & 0 \ \, 2 \ \, 0  & 0 \ \, 2 \ \, 0 \\
-2.158744056768  & 0 \ \ 1 &  0 \ \ 1  & 1 \ \, 0 \ \, 0  & 1 \ \, 0 \ \, 0 \\
 \hline
 \end{tabular}
 \caption{Spectra of various transfer matrices with
 $Q=3$, subject to periodic ($I$) or different twisted periodic ($Z_2$ and
 $Z_3$) boundary conditions, as defined in the text. The first column gives
 the free energies $f_i=-L^{-1}\log(\Lambda_i)$, here for width $L=2$ and
 temperature variable $x=5$. Subsequent columns give the multiplicity of each
 $f_i$.}
 \label{tab1}
\end{table}

Due to the rules of the RSOS model, the heights living on the direct and dual
lattices have opposite parities. The transfer matrix can therefore be decomposed
in two sectors,
$T_{\rm RSOS} = T_{\rm RSOS}^{\rm even} \oplus T_{\rm RSOS}^{\rm odd}$,
henceforth referred to as even and odd. In the even sector, direct heights
take {\em odd} values and dual heights {\em even} values (and vice versa for
the odd sector).%
\footnote{This definition of parity may appear somewhat strange; it is
motivated by the fact that it implies
$T^{\rm even}_{\rm RSOS}(x)=T_{\rm spin}(x)$ for the Ising case $p=4$,
cf.~Eq.~(\ref{IsingTM}).}
Results with standard (i.e., untwisted) periodic boundary
conditions in the vertical direction are given in the columns labeled
$I$ in Table~\ref{tab1}.

In the spin representation $T_{\rm spin}(x)$ has been defined above (recall that
$x=Q^{-1/2}({\rm e}^K-1)$). We also introduce a related transfer matrix
\beq
 T_{\rm spin}^{\rm dual}(x) \equiv x^{2L} T_{\rm spin}(x^{-1}),
 \label{Tspindual}
\eeq
as well as the corresponding
partition function $Z_{\rm spin}^{\rm dual}(x) \equiv
{\rm Tr} \, (T_{\rm spin}^{\rm dual}(x))^N$. The appearance of the dual
temperature, $x_{\rm dual} = x^{-1}$ explains the terminology. More
precisely, on a planar lattice one has the fundamental duality relation
\cite{Wu82}
\beq
 Z_{\rm spin}(x) = Q^{-1} x^E \tilde{Z}_{\rm spin}(x^{-1}),
 \label{funddual}
\eeq
where $\tilde{Z}$ must be evaluated on the dual lattice. For a square lattice
with toroidal boundary conditions, the dual and direct lattices are isomorphic,
but Eq.~(\ref{funddual}) breaks down because of effects of non-planarity.

Referring to Table~\ref{tab1}, we observe that the leading eigenvalues of the
four transfer matrices introduced this far (i.e., $T_{\rm RSOS}^{\rm even}$,
$T_{\rm RSOS}^{\rm odd}$, $T_{\rm spin}$ and $T_{\rm spin}^{\rm dual}$) all
coincide. On the other hand, for any two $T$ chosen among these four, some of
the sub-leading eigenvalues coincide, whilst others are different.
So the discrepancy between the four corresponding partition functions appears to
be a boundary effect which vanishes in limit $N\to\infty$.
However, when taking differences of the multiplicities we discover a surprising
relation:
\beq
 2(Z^{\rm even}_{\rm RSOS}(x)-Z^{\rm odd}_{\rm RSOS}(x)) =
 Z_{\rm spin}(x)-Z^{\rm dual}_{\rm spin}(x).
 \label{relQ3}
\eeq
Note that the leading eigenvalues cancel on both sides of this relation.

At the selfdual point $x=1$, we find that the spectra of
$T_{\rm RSOS}^{\rm even}$ and $T_{\rm RSOS}^{\rm odd}$ coincide completely,
as do those of $T_{\rm spin}$ and $T_{\rm spin}^{\rm dual}$. In particular,
both sides of Eq.~(\ref{relQ3}) vanish. It is however still true that
subleading eigenvalues of $T_{\rm RSOS}^{\rm even}$ and $T_{\rm spin}$ differ.

More relations can be discovered by introducing twisted periodic boundary
conditions in the transfer matrices. For the RSOS model this can be done by
twisting the heights, $h \to p-h$, when traversing a horizontal seam. Note
that this transformation makes sense at it leaves the weights of
Eq.~(\ref{weights}) invariant, since $S_h = S_{p-h}$. The shape of the seam
can be deformed locally without changing the corresponding partition function;
we can thus state more correctly that the seam must be homotopic to the
horizontal principal cycle of the torus. Note also that the twist is only well
defined for even $p$ (and in particular for $Q=3$, $p=6$), since the heights
on the direct and dual lattice must have fixed and opposite parities in order
to satisfy the RSOS constraint. In Table~\ref{tab1}, this twist is labeled
$Z_2$, since it amounts to exploiting the $Z_2$ symmetry of the underlying
Dynkin diagram $A_{p-1}$. Comparing again multiplicities we discover a second
relation
\beq
 Z^{\rm even}_{\rm RSOS}(x)-Z^{\rm odd}_{\rm RSOS}(x) =
 Z^{{\rm even},Z_2}_{\rm RSOS}(x)-Z^{{\rm odd},Z_2}_{\rm RSOS}(x).
 \label{twQ3}
\eeq

The relations (\ref{relQ3}) and (\ref{twQ3}) are special cases of
relations that hold for all $x$, $L$ and $N$, and for all even $p$.
The general relations (see Eqs.~(\ref{maindual}) and (\ref{diffZ2})) are
stated and proved in section~\ref{sec3} below.

In the particular case of $Q=3$ one can define two different ways of twisting
the spin representation, which will lead to further relations. The first
type of twist shall be referred to as a $Z_2$ twist, and consist in interchanging
spin states $S_i=1$ and $S_i=2$ across a horizontal seam, whereas the spin
state $S_i=3$ transforms trivially. The second type of twist, the
$Z_3$ twist, consists in permuting the three spin states cyclically when
traversing a horizontal seam. The spectra of the corresponding transfer
matrices are given in Table~\ref{tab1}.

This leads to another relation between the spectra in the spin
representation (stated here in terms of the corresponding partition
functions)
\beq
 Z^{Z_2}_{\rm spin}(x) = Z^{{\rm dual},Z_2}_{\rm spin}(x),
 \label{duZ2}
\eeq
as well a two further relations linking the spin and RSOS representations:
\bea
 Z^{\rm even}_{\rm RSOS}(x)-Z^{\rm odd}_{\rm RSOS}(x) &=&
 -\left( Z^{Z_3}_{\rm spin}(x)-Z^{{\rm dual},Z_3}_{\rm spin}(x) \right),
 \label{duZ3} \\
 Z^{\rm even}_{\rm RSOS}(x) + Z^{{\rm even},Z_2}_{\rm RSOS}(x) &=&
 Z_{\rm spin}(x) + Z^{Z_2}_{\rm spin}(x). \label{comboZ2}
\eea
These relations are proved in section~\ref{secQ3} below.

\section{Relations between partition functions}
\label{sec3}

\subsection{Weights in the cluster and RSOS pictures}

\begin{figure}
\begin{center}
 \leavevmode
 \epsfysize=60mm{\epsffile{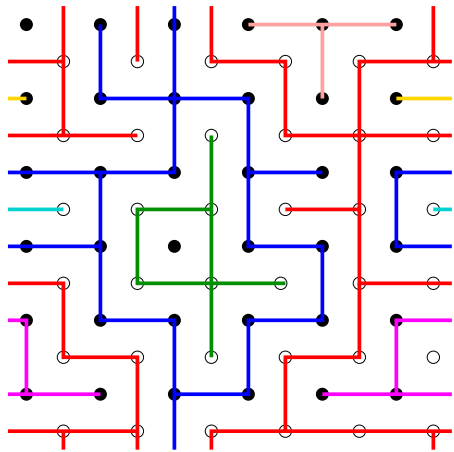}}
 \end{center}
 \protect\caption[3]{A possible cluster configuration on a $6 \times 6$ torus.
Direct and dual vertices are shown as filled and empty black circles.
Clusters (direct or dual), other than isolated vertices, are depicted here
using distinct colours, for convenience in appreciating the periodic
boundary conditions. There are six direct clusters and four dual clusters.
One direct cluster and one dual cluster are non-homotopic to a point.}
 \label{fig2}
\end{figure}

A possible configuration of clusters on a $6 \times 6$ torus is shown in
Fig.~\ref{fig2}. It can be thought of as a random tessellation using the
two tiles of Fig.~\ref{fig1}. For simplicity we show here only the clusters
(direct or dual) and not their separating boundaries (given by the thin red
lines in Fig.~\ref{fig1}). Two clusters having a common boundary are said to
be neighbouring. For convenience in visualising the periodic boundary conditions
the thick lines depicting the clusters have been drawn using different colours
(apart from clusters consisting of just one isolated vertex, which are
all black).

To compute the contribution of this configuration to $Z_{\rm cluster}(x)$,
each direct cluster is weighed by a factor of $Q$, and each coloured direct
edge carries a factor of $v=Q^{1/2}x$. Note in particular that the cluster
representation does not distinguish clusters of non-trivial topology (i.e.,
clusters which are not homotopic to a point). In the following we shall
call such clusters non-trivial; clusters which are homotopic to a point are
then referred to as trivial.

\begin{figure}
\begin{center}
 \leavevmode
 \epsfysize=20mm{\epsffile{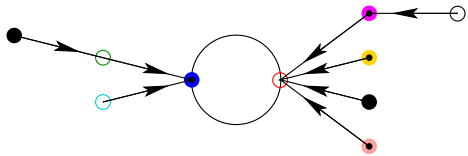}}
 \end{center}
 \protect\caption[3]{Pasquier graph corresponding to the cluster configuration
of Fig.~\ref{fig2}. Direct (resp.\ dual) clusters are shown as filled
(resp.\ empty) circles, using the same colour coding as in Fig.~\ref{fig2}.
Neighbouring clusters are connected by an edge. The arrows are explained
in the text.}
 \label{fig3}
\end{figure}

The contribution of this same configuration to $Z_{\rm RSOS}$ consists of
\be
 \item a global factor of $Q^{V/2}$ coming from the prefactor
 of Eq.~(\ref{TM}),
 \item a factor of $x$ for each coloured direct edge \cite{SB}, and
 \item an $x$-independent factor due to the topology of the (direct and dual)
       clusters \cite{Pasquier}.
\ee
The interest is clearly concentrated on this latter, topological factor which we denote
$w$ in the following.
For a given
cluster configuration its value can be computed from the adjacency information
of the clusters. This information is conveniently expressed in the form of a
Pasquier graph \cite{Pasquier}; for the cluster configuration of
Fig.~\ref{fig2} this graph is shown in Fig.~\ref{fig3}.

The rules for drawing the Pasquier graph in the general case are as
follows.  Each cluster is represented by a vertex, and vertices
representing neighbouring clusters (i.e., clusters having a common
boundary) are joined by a directed edge. An edge directed from vertex
$A$ to vertex $B$ means that the common boundary is surrounding
cluster $A$ and is surrounded by cluster $B$.  In particular, the
in-degree $b_{\rm in}$ of a cluster is the number of boundaries
surrounded by that cluster, and the out-degree $b_{\rm out}$ is the
number of boundaries surrounding the cluster. By definition,
a boundary separating a non-trivial cluster from a trivial one is
said to be surrounded by the non-trivial cluster; note that there is
necessarily at least one non-trivial cluster. We do not assign any
orientation to edges joining two non-trivial clusters,
since in that case the notion of surrounding is nonsensical.

The topological structure of the Pasquier graph is characterised by
the following three properties:
\be
 \item The graph is bicolourable, with one colour (represented by filled
 circles in Fig.~\ref{fig3}) corresponding to direct clusters and the
 other (empty circles in Fig.~\ref{fig3}) to dual clusters.
 \item The vertices corresponding to non-trivial clusters and the
 undirected edges form a cycle.
 \item Each vertex corresponding to a non-trivial cluster is the root
 of a (possible empty) tree, whose vertices correspond to trivial clusters.
 The edges in the tree are directed towards the root.
\ee
Property 3 is easily proved by noticing that vertices corresponding to trivial
clusters have all $b_{\rm out}=1$, i.e., such clusters have a unique external
boundary. Regarding property 2, we shall define the order $n$ of the Pasquier
graph as the number of undirected edges. By property 1, $n$ is even. When
$n=0$ we shall call the graph degenerate; this corresponds to a situation
in which a single cluster (direct or dual) is non-trivial. 

In the RSOS picture, each configuration of the clusters (such as the one on
Fig.~\ref{fig2}) corresponds to many different height configurations. The
topological ($x$-independent) contribution to the weight of a cluster
configuration in $Z_{\rm RSOS}$ is therefore obtained by summing weights in
the RSOS model with $x=1$ over all height configurations which are compatible
with the given cluster configuration \cite{Pasquier}. This contribution can
be computed from the Pasquier graph by using the incidence matrix $G_{p-1}$
of the Dynkin diagram $A_{p-1}$, as we now review.

Let $w$ be the weight of a given Pasquier graph, and let $w'$ be the weight of
the graph in which a leaf of one of its trees (as well as its adjacent
outgoing edge) has been removed. More precisely, $w$ is the weight of a
cluster configuration with given heights on each cluster, and $w'$ is the
partial sum of such weights over all possible heights of the leaf cluster. Let
$j$ be the height of the leaf, and let $i$ be the height of its parent. Then
\beq
 w = w' (S_i)^{-1} \sum_{1 \le j \le p-1} (G_{p-1})_{ij} S_j
   = w' (S_i)^{-1} Q^{1/2} S_i = Q^{1/2} w',
 \label{trivial}
\eeq
where in the first equality we used that the weight of a cluster at
height $h$ is $S_h^{b_{\rm out}-b_{\rm in}}$ \cite{Pasquier}, and in
the second that $\{S_j\}$ is an eigenvector of $G_{p-1}$ with eigenvalue
$Q^{1/2}$. Iterating the argument until all the trees of the Pasquier
graph have been removed, we conclude that each trivial cluster carries
the weight $Q^{1/2}$.

We have then
\beq
 w = Q^{(l-n)/2} w_{\rm c},
 \label{onlyw}
\eeq
where $w_{\rm c}$ is the weight of the cycle of the Pasquier graph.
It corresponds to the number of closed paths of length $n$ on the Dynkin
diagram
\beq
 w_{\rm c} = {\rm Tr} \, (G_{p-1})^n = \sum_{1 \le k \le p-1}
 \left( 2 \cos(k \pi / p) \right)^n,
 \label{wc}
\eeq
where we have used the eigenvalues of $G_{p-1}$. Note that, in contrast to the
case of trivial clusters, {\em all} the eigenvalues contribute to the combined
weight $w_{\rm c}$ of the non-trivial clusters, and that this weight cannot in
general be interpreted as a product of individual cluster weights. (We also
remark that it is not a priori obvious that the right-hand side of
Eq.~(\ref{wc}) is an integer.)

\subsection{Coincidence of highest eigenvalues}
\label{highest}

As an application we now argue that the dominant eigenvalues of the transfer
matrices $T_{\rm RSOS}$, $T_{\rm cluster}$, $T_{\rm spin}$ coincide for any
width $L$. We suppose $x>0$ so that all weights are positive; this guarantees
in particular that standard probabilistic arguments apply.

Consider first $T_{\rm cluster}$ and $T_{\rm spin}$, supposing $Q$ a
positive integer. Since the system is quasi one-dimensional, with $L
\ll N$, configurations having clusters of linear extent much larger
than $L$ are exceedingly rare and can be neglected. In particular,
almost surely no cluster will wrap around the system in the horizontal
direction. Writing $Z_{\rm cluster} \sim (\Lambda_{\rm c})^N$ and
$Z_{\rm spin} \sim (\Lambda_{\rm s})^N$, the choice of boundary
conditions in the horizontal direction will thus have no effect on the
values of $\Lambda_{\rm c}$ and $\Lambda_{\rm s}$. We therefore switch
to free boundary conditions in the horizontal direction. Then it is
possible \cite{Salas01} to write $Z_{\rm cluster} = \langle f |
(T_{\rm cluster})^N | i \rangle$ for suitable initial and final
vectors $| i \rangle$ and $\langle f |$. It is not difficult to see,
using the Perron-Frobenius theorem, that these vectors both contain a
non-vanishing component of the dominant eigenvector of $T_{\rm
  cluster}$. We conclude that $\Lambda_{\rm c}$ must be the dominant
eigenvalue of $T_{\rm cluster}$. Likewise, $\Lambda_{\rm s}$ is the
dominant eigenvalue of $T_{\rm spin}$. The conclusion follows by
noting that $Z_{\rm cluster}=Z_{\rm spin}$ by construction.

We now turn to $T_{\rm RSOS}$ and $T_{\rm cluster}$, supposing
$Q^{1/2}=2\cos(\pi/p)$, cf.~Eq.~(\ref{rootunity}). As before we
impose free horizontal boundary conditions on the cluster model. Then,
since the resulting lattice is {\em planar}, we recall
that $Z_{\rm cluster}$ can be written as well in terms of the boundaries
separating direct and dual clusters \cite{BKW,BaxterBook}
\beq
 Z_{\rm cluster} = Q^{V/2} \sum_{\cal C} x^{e({\cal C})} Q^{l({\cal C})/2},
 \label{Zloop}
\eeq
where the configurations ${\cal C}$ correspond bijectively to those of
Eq.~(\ref{ZPotts}), and $l({\cal C})$ is the number of cluster boundaries
(loops). For $N\to\infty$ we have $Z_{\rm RSOS} \sim (\Lambda_{\rm r})^N$, and
to conclude that $\Lambda_{\rm r} = \Lambda_{\rm c}$ it suffices to show that
asymptotically $Z_{\rm RSOS} \sim Z_{\rm cluster}$. Consider now a typical
cluster configuration. Almost surely, the corresponding Pasquier diagram will
be of order $n \sim N$, and in particular $n \gg 1$. Hence $w_{\rm c} \sim
Q^{n/2}$ from Eq.~(\ref{wc}). It follows that also in the RSOS picture each
cluster boundary carries the weight $Q^{1/2}$, regardless of its homotopy. The
conclusion follows.

In section~\ref{sec2} we have introduced the decomposition of the RSOS model
into even and odd subsectors. In the even sector, heights on direct
(resp.\ dual) clusters are {\em odd} (resp.\ {\em even}). In particular,
we have $Z_{\rm RSOS} = Z_{\rm RSOS}^{\rm even}+Z_{\rm RSOS}^{\rm odd}$.
Note that since that the largest and smallest eigenvalue in Eq.~(\ref{wc})
differ just by a sign change, we have the slightly more precise statement
for $N\to\infty$:
\beq
 Z_{\rm RSOS} \simeq 2 Z_{\rm RSOS}^{\rm even}
 \simeq 2 Z_{\rm RSOS}^{\rm odd} \simeq 2 Z_{\rm cluster}.
\eeq
In particular, the largest eigenvalues of $T_{\rm RSOS}^{\rm even}$
and $T_{\rm RSOS}^{\rm odd}$ coincide, in agreement with the numerical
data of Table~\ref{tab1}.

\subsection{Duality relation for
$Z_{\rm RSOS}^{\rm even}-Z_{\rm RSOS}^{\rm odd}$}

We now compare the contributions to the partition functions
$Z_{\rm RSOS}^{\rm even}$ and $Z_{\rm RSOS}^{\rm odd}$ for a
given cluster configuration (summed over all possible heights
assignments with the specified parity). The argument that trivial
clusters carry a weight $Q^{1/2}$ is unchanged,
cf.~Eq.~(\ref{trivial}), and holds irrespective of parity. We can thus
limit the discussion to the weight $w_{\rm c}$ of the cycle in the
Pasquier graph.

We first show that
\beq
 w_{\rm c}^{\rm even} = w_{\rm c}^{\rm odd} = \frac12 {\rm Tr} \, (G_{p-1})^n
 \label{halftrace}
\eeq
for non-degenerate Pasquier graphs (i.e., of order $n \neq 0$). In this
non-degenerate case, the numbers of direct and dual non-trivial clusters
are equal, whence $n=2k$ is even. By definition $w_{\rm c}^{\rm even}$ is
the number of height assignments $\{h_1,h_2,\ldots,h_{2k}\}$ such that
$h_i = 1,2,\ldots,p-1$ and $|h_{i+1}-h_i| = 1$ (we consider $i$ modulo $2k$),
with $h_1$ even. Now by a cyclic relabeling, $i \to i+1$ (mod $2k$), each
such height assignment is mapped bijectively to a height assignment in which
$h_1$ is odd. It follows that $w_{\rm c}^{\rm even} = w_{\rm c}^{\rm odd}$.%
\footnote{It is amusing to rephrase the result (\ref{halftrace}) in
  terms of Dynkin diagrams. Let $n_i$ be the number of closed paths of
  length $2k$ on $A_{p-1}$, starting and ending at site $i$. Then
  Eq.~(\ref{halftrace}) amounts to $\sum_{i \ {\rm even}}n_i = \sum_{i
    \ {\rm odd}}n_i$. For $p$ odd this is obvious, since $n_i =
  n_{p-i}$ by the $Z_2$ symmetry of $A_{p-1}$; however for $p$ even
  this is a non-trivial statement (though straightforward to prove,
  using generating function techniques for example).}

Consider now the degenerate case $n=0$ with just a single non-trivial cluster
(which will then span both periodic directions of the torus). Then, counting
just the number of available heights of a given parity, the contribution of the
cycle to respectively $Z_{\rm RSOS}^{\rm even}$ and $Z_{\rm
RSOS}^{\rm odd}$ read ($\lfloor x \rfloor$ denotes the integer part of $x$)
\beq
 w_{\rm c}^{\rm even}  = \lfloor p/2 \rfloor, \qquad
 w_{\rm c}^{\rm odd} = \lfloor (p-1)/2 \rfloor,
 \label{intpart}
\eeq
if the non-trivial cluster is direct (if it is dual, permute the labels
even and odd). 
In particular, $w_{\rm c}^{\rm even}=w_{\rm c}^{\rm odd}$ for $p$ odd and we
deduce that
\beq
 Z_{\rm RSOS}^{\rm even}(x) = Z_{\rm RSOS}^{\rm odd}(x) \mbox{ for $p$ odd}.
 \label{evenodd}
\eeq

Of course, Eq.~(\ref{evenodd}) can be proved in a much more elementary
way by noticing that for $p$ odd the RSOS model is symmetric under the
transformation $h \to p-h$, which exchanges the parity of the heights.
This even implies the stronger statement $T_{\rm RSOS}^{\rm even} =
T_{\rm RSOS}^{\rm odd}$.  On the other hand, for $p$ even, the
transformation $h \to p-h$ does not change the parity, and $T_{\rm
  RSOS}^{\rm even} \neq T_{\rm RSOS}^{\rm odd}$; the two matrices do
not even have the same dimension.

The purpose of presenting the longer argument leading to Eq.~(\ref{evenodd})
is to make manifest that this relation breaks down for even $p$ exactly
because of configurations represented by degenerate Pasquier graphs.
However a weaker relation holds true for any parity of $p$:
\beq
 Z_{\rm RSOS}^{\rm even}(x) = x^E Z_{\rm RSOS}^{\rm odd}(x^{-1}).
 \label{anyparity}
\eeq
Note that it implies, as a corollary, that Eq.~(\ref{evenodd}) also
holds for $p$ even provided that $x=1$.

To prove Eq.~(\ref{anyparity}) we again argue configuration by
configuration. Each cluster configuration is in bijection with a
``shifted'' configuration obtained by keeping fixed the coloured edges
and moving the whole lattice by half a lattice spacing in both
directions (or equivalently, exchanging the direct and dual vertices).
The shifted configuration has the same Pasquier graph as the original
one, except for an exchange of direct and dual vertices and thus of
the parity of the heights on direct vertices.  We conclude that
$w_{\rm c}^{\rm even}$ computed for the original configuration equals
$w_{\rm c}^{\rm odd}$ computed for the shifted configuration, and vice
versa. This implies Eq.~(\ref{anyparity}) upon noticing that the
factors of $x$ correct the weighing of the coloured direct edges (we
have used that the sum of direct and dual coloured edges equals $E$).

Subtracting Eq.~(\ref{anyparity}) from the relation obtained from
Eq.~(\ref{anyparity}) under $x \to x^{-1}$ gives a duality
relation for $Z_{\rm RSOS}^{\rm even}-Z_{\rm RSOS}^{\rm odd}$:
\beq
 Z_{\rm RSOS}^{\rm even}(x)-Z_{\rm RSOS}^{\rm odd}(x) =
 -x^E \left( Z_{\rm RSOS}^{\rm even}(x^{-1})-Z_{\rm RSOS}^{\rm odd}(x^{-1})
 \right).
\eeq
We shall now see that the left-hand side of this relation can be related
to a difference of partition functions in the cluster picture.

\subsection{A relation between RSOS and cluster partition functions}

We have already mentioned above, in Eq.~(\ref{Zloop}), that for a {\em planar}
lattice $Z_{\rm cluster}$ can be written in terms of the boundaries (loops)
that separate direct and dual clusters \cite{BKW,BaxterBook}
\beq
 Z_{\rm cluster} = Q^{V/2} \sum_{\cal C} x^{e({\cal C})} Q^{l({\cal C})/2},
 \label{Zloop2}
\eeq
where $l({\cal C})$ is the number of cluster boundaries.
This result is obtained from Eq.~(\ref{ZPotts}) by using the Euler relation
for a planar graph, $l({\cal C}) + V = 2 c({\cal C}) + e({\cal C})$.

On a torus, things are slightly more complicated. The Euler relation must be
replaced by
\bea
 2 + l({\cal C}) + V &=& 2 c({\cal C}) + e({\cal C})
 \mbox{ if a direct cluster spans both periodic directions}, \nn \\
 l({\cal C}) + V &=& 2 c({\cal C}) + e({\cal C}) \mbox{ otherwise}.
 \label{Euler}
\eea
To prove Eq.~(\ref{Euler}) we proceed by induction. Initially, when ${\cal C}$
is the state with no coloured direct edge, we have
$l({\cal C})=c({\cal C})=V$ and $e({\cal C})=0$, whence the second of the
relations indeed holds true. Any other configuration ${\cal C}$ can
be obtained from the initial one by successively colouring direct
edges (and uncolouring the corresponding dual edges). When colouring a
further direct edge, there are several possibilities:
\be
 \item The edge joins two clusters which were formerly distinct. The
 changes in the parameters of Eq.~(\ref{Euler}) are then
 $\Delta l = -1$ (the outer boundaries of the two clusters join to form
 the outer boundary of the amalgamated cluster), $\Delta c=-1$ and
 $\Delta e = 1$. Thus, the changes to the left- and right-hand sides
 of Eq.~(\ref{Euler}) cancel out.
 \item The edge joins two vertices which were already in the same cluster.
 Then $\Delta l = 1$ (the operation creates a cycle in the cluster which
 must then acquire an inner boundary), $\Delta c=0$ and $\Delta e = 1$. This
 again  maintains Eq.~(\ref{Euler}). The same changes are valid when the added
 edge makes the cluster wrap around the {\em first} of the two periodic
 directions: no inner boundary is created in this case, but the cluster's
 outer boundary breaks into two disjoint pieces.
 \item The edge makes, for the first time, the cluster wrap around {\em both}
 periodic directions. Then $\Delta l = -1$ (the two outer boundaries
 coalesce), $\Delta c=0$ and $\Delta e=1$. Thus one jumps from the second
 to the first of the relations (\ref{Euler}).
\ee

We conclude that on a torus, Eq.~(\ref{Zloop2}) must be replaced by
\beq
 Z_{\rm cluster} = Q^{V/2} \sum_{\cal C} x^{e({\cal C})}
 Q^{l({\cal C})/2+\eta({\cal C})},
 \label{Zloop3}
\eeq
where, in the language of Pasquier graphs, $\eta({\cal C})=1$ if $n=0$
and the non-trivial cluster is {\em direct}, and $\eta({\cal C})=0$ in
all other cases. Note that $n$ is the number of non-trivial cluster
boundaries and $l-n$ the number of trivial boundaries.

Eq.~(\ref{Zloop3}) can be used to prove the following relation between
RSOS and cluster partition functions:
\beq
 (Q-1) \left( Z_{\rm RSOS}^{\rm even}(x)-Z_{\rm RSOS}^{\rm odd}(x) \right) =
 Z_{\rm cluster}(x) - x^E Z_{\rm cluster}(x^{-1}) \mbox{ for $p$ even}.
 \label{maindual}
\eeq
Note that we do not claim the validity of Eq.~(\ref{maindual}) for $p$ odd,
since then the left-hand side vanishes by Eq.~(\ref{evenodd}) whereas the
right-hand side is in general non-zero. We also remark that for $p=6$,
Eq.~(\ref{maindual}) reduces to Eq.~(\ref{relQ3}) which was conjectured
based on numerical evidence in section~\ref{sec2}.

We now prove Eq.~(\ref{maindual}) by showing the each cluster configuration
gives equal contributions to the left- and the right-hand sides. When
evaluating the second term on the right-hand side we consider instead
the shifted configuration. This ensures that the contribution to all terms
in Eq.~(\ref{maindual}) yields the same power of $x$; we therefore
only the topological weight $w$ (cf.~Eq.~(\ref{onlyw}) in the following argument.

The contribution of non-degenerate Pasquier graphs to the left-hand side
of Eq.~(\ref{maindual}) is zero by Eq.~(\ref{halftrace}) and the discussion
preceding it. The contribution of such graphs to the right-hand side
also vanishes, since the original and shifted configurations have the same
number of cluster boundaries $l$, and in both cases $\eta = 0$ in
Eq.~(\ref{Zloop3}).

Consider next the contribution of a degenerate Pasquier graph where
the non-trivial cluster is {\em direct}. The contribution to the
left-hand side of Eq.~(\ref{maindual}) is $(Q-1) Q^{l/2}$, since
from Eq.~(\ref{intpart}) $w_{\rm c}^{\rm even}-w_{\rm c}^{\rm odd}=1$.
As to the right-hand side, note that for the first term, $Z_{\rm cluster}(x)$,
we have $\eta=1$ in Eq.~(\ref{Zloop3}), whereas for the second term,
$x^E Z_{\rm cluster}(x^{-1})$, we use the shifted configuration as announced,
whence $\eta=0$. The total contribution to the right-hand side of
Eq.~(\ref{maindual}) is then $Q^{l/2+1}-Q^{l/2}$ as required.

When the non-trivial cluster is {\em dual} a similar argument can be given
(there is a sign change on both sides); Eq.~(\ref{maindual}) has thus been
proved.

Note that while Eq.~(\ref{maindual}) itself reduces to a tautology at the
selfdual point $x=1$, one can still obtain a non-trivial relation by
taking derivatives with respect to $x$ on both sides before setting $x=1$.
For example, deriving once one obtains for $x=1$:
\beq
\langle e \rangle_{\rm RSOS}^{\rm even} - \langle e \rangle_{\rm RSOS}^{\rm odd}= 
2 Z_{\rm cluster} ((Q-1) Z_{\rm RSOS})^{-1}
(\langle e \rangle_{\rm cluster} - \langle e_{\rm dual} \rangle_{\rm cluster})
 \label{derividentity}
\eeq
where $e_{\rm dual}=E-e$ is the number of coloured dual edges. Higher derivatives give relations 
involving higher moments of $e$ and $e_{\rm dual}$. Eq.~(\ref{anyparity}) gives similar relations 
using the same procedure, for example:
\beq
\langle e \rangle_{\rm RSOS}^{\rm even} = \langle e_{\rm dual} \rangle_{\rm RSOS}^{\rm odd}
\eeq
which can be proved directly considering shifted configurations.

\subsection{Topology of the non-trivial clusters}

In the following sections we consider twisted models, and it is necessary
to be more careful concerning the topology of the non-trivial clusters.

Consider first the non-degenerate case, $n \neq 0$. Each of the boundaries
separating two non-trivial clusters takes the form of a non-trivial, non-self
intersecting loop on the torus. Assign to each of these loops an arbitrary
orientation. The homotopy class of an oriented loop is then characterised by a
pair of integers $(i_1,i_2)$, where $i_1$ (resp.\ $i_2$) indicates how many
times the horizontal (resp.\ vertical) principal cycle of the torus is crossed
in the positive direction upon traversing the oriented loop once. We recall a
result \cite{FSZ} stating that 1) $|i_1|$ and $|i_2|$ are coprime (in
particular they have opposite parities), and 2) the {\em relative}
orientations of the non-trivial loops defined by a given cluster configuration
can be chosen so that all the loops have the same $(i_1,i_2)$. Further, by a
{\em global} choice of orientations, we can suppose that $i_1 \ge 0$. The sign
of $i_2$ is then changed by taking an appropriate mirror image of the
configuration; since this does not affect the weights in the cluster and RSOS
models we shall henceforth suppose that $i_2 \ge 0$ as well.

By an abuse of language, we shall define the homotopy class of the non-trivial
clusters by the same indices $(i_1,i_2)$ that characterise the non-trivial
loops. For example, clusters percolating only horizontally correspond to
homotopy class $(1,0)$, and clusters percolating only vertically correspond to
class $(0,1)$. Note that there are more complicated clusters which have both
$i_1>0$ and $i_2>0$, and that if one of the indices is $\ge 2$ the other must
be $\ge 1$.

Finally, in the degenerate case $n=0$, all the loops surrounded by the
non-trivial cluster are actually trivial. The homotopy class of the cluster is
then defined to be $(i_1,i_2)=(0,0)$.

\subsection{Twisted RSOS model}

For even $p$, the RSOS model can be twisted by imposing the identification
$h \to p-h$ upon crossing a horizontal seam, as already explained before
Eq.~(\ref{twQ3}). We refer to this as $Z_2$ type twisted boundary conditions.

Those new boundary conditions change the weights of the Pasquier graphs.
The trivial clusters still have weight $Q^{1/2}$ (the seam can be locally
deformed so as to avoid traversing these clusters), whereas the weight of
the cycle $w_{\rm c}$ is modified.

Consider first the non-degenerate case $n\neq 0$. If a non-trivial
cluster has $i_2$ odd (where $i_2$ corresponds to the direction perpendicular to the
seam) its height $h$ is fixed by $h=p-h$, whence $h=p/2$. But since
$n \ge 2$, there must be both a direct and a dual cluster wrapping in
this way, and since their heights have opposite parities they cannot
both equal $p/2$. So such a configuration is incompatible with the $Z_2$
boundary conditions. 

Suppose instead that $n>0$ clusters (i.e., necessarily $n/2$ direct
and $n/2$ dual) have $i_2$ even.  
The weight $w_{\rm c}$ 
is no longer given by Eq.~(\ref{wc}), but rather by
\beq
 w_{\rm c}^{Z_2} = {\rm Tr} \, \left[ (G_{p-1})^n J_{p-1} \right],
\eeq
where the matrix
\beq
 J_{p-1} = \left(
 \begin{tabular}{ccccc}
 0        & $\cdots$ & $\cdots$ & 0        & 1         \vspace{-0.2cm} \\
 $\vdots$ &          &          & 1        & 0         \vspace{-0.2cm} \\
 $\vdots$ &          & /        &          & $\vdots$  \vspace{-0.2cm} \\
 0        & 1        &          &          & $\vdots$  \vspace{0.00cm} \\
 1        & 0        & $\cdots$ & $\cdots$ & 0         \\
 \end{tabular} \right)
\eeq
of dimension $p-1$ implements the jump in height $h \to p-h$ due to
the seam. It is easy to see that the matrices
$G_{p-1}$ and $J_{p-1}$ commute (physically this is linked to the fact that
the cut can be deformed locally) and thus have the same eigenvectors.
These are of the form $|v_k \rangle = \left \{ \sin (\pi k h/p) \right
\}_{h=1,2,\ldots,p-1}$.  The corresponding eigenvalues of $G_{p-1}$
read $\lambda_k = 2 \cos (\pi k /p)$ for $k=1,2,\ldots,p-1$. Since the
eigenvectors with $k$ odd (resp.\ even) are symmetric (resp.\
antisymmetric) under the transformation $h \to p-h$, the
eigenvalues of $J_{p-1}$ are $(-1)^{k+1}$. We conclude that Eq.~(\ref{wc})
must be replaced by
\beq
 w_{\rm c}^{Z_2} = \sum_{1 \le k \le p-1} (-1)^{k+1}
 \left( 2 \cos(k \pi/p) \right)^n \mbox{ for $n \neq 0$}.
 \label{wctw}
\eeq
For the degenerate case $n=0$, one has simply $w_{\rm c}^{Z_2}=1$, since
the height of the non-trivial cluster is fixed to $p/2$.

As in the untwisted sector we can impose given parities on the direct
and dual clusters. This does not change the weighing of trivial
clusters. For non-trivial clusters with $n \neq 0$ we have $w_{\rm
  c}^{{\rm even},Z_2}=w_{\rm c}^{{\rm odd},Z_2}= \frac12 w_{\rm
c}^{Z_2}$ for the same reasons as in Eq.~(\ref{halftrace}), but with
$w_{\rm c}^{Z_2}$ now given by Eq.~(\ref{wctw}). Finally, for $n=0$
one finds for a direct non-trivial cluster
\beq
 w_{\rm c}^{{\rm even},Z_2} = p/2 \mbox{ mod } 2, \qquad
 w_{\rm c}^{{\rm odd},Z_2} = 1-p/2 \mbox{ mod } 2.
 \label{wcn0}
\eeq
For a dual non-trivial cluster, exchange the labels even and odd.

We have the following relation between the twisted and untwisted
RSOS models:
\beq 
 Z_{\rm RSOS}^{\rm even}(x)-Z_{\rm RSOS}^{\rm odd}(x) =
 (-1)^{p/2+1} \left(
 Z_{\rm RSOS}^{{\rm even},Z_2}(x)-Z_{\rm RSOS}^{{\rm odd},Z_2}(x) \right)
 \mbox{ for $p$ even}.
 \label{diffZ2}
\eeq
Indeed, the two sides are non-zero only because of parity effects
in $w_{\rm c}$ when $n=0$. The relation then follows by comparing
Eqs.~(\ref{intpart}) and (\ref{wcn0}). Note that Eq.~(\ref{diffZ2})
generalises Eq.~(\ref{twQ3}) which was conjectured
based on numerical evidence in section~\ref{sec2}.

\subsection{Highest eigenvalue of the twisted RSOS model}
\label{hieigtw}

We now argue that the dominant eigenvalue of $T_{\rm RSOS}^{Z_2}(x)$ coincides
with a subdominant eigenvalue of $T_{\rm RSOS}(x)$ for any width $L$.
We consider first the case of $p/2$ odd.

The configurations contributing to $Z_{\rm RSOS}^{Z_2}(x)$ are those in which
non-trivial clusters have $i_2$ even, which includes in particular the
degenerate case (the non-trivial cluster being direct or dual depending on the
parity considered). In the limit where $L \ll N$, the typical configurations
correspond therefore to degenerate configurations, with the non-trivial
cluster being direct in the even sector and dual in the odd sector, see
Eq.~(\ref{wcn0}). Because of these parity effects, the dominant eigenvalues
are not the same for both parities (except of course for $x=1$).

For such degenerate configurations, the weights corresponding to the twisted
and untwisted models are different, but since the difference $w_c^{\rm
even}-w_c^{\rm odd}$ is the same independently of the twist, we have
Eq.~(\ref{diffZ2}). Since the dominant eigenvalues do not cancel from the
right-hand side of that relation, they also contribute to the left-hand side.

We can therefore write, in each parity sector, $Z_{\rm RSOS}(x) \sim Z_{\rm
RSOS}^{Z_2}(x) + Z_{\rm RSOS}^0(x)$, where $Z_{\rm RSOS}^0(x)$ accounts for
configurations in which no cluster percolates horizontally (there will
therefore be at least one, and in fact almost surely many, clusters
percolating vertically). If this had been an exact identity, we could resolve
on eigenvalues of the corresponding transfer matrices and conclude that the
eigenvalues of $T_{\rm RSOS}^{Z_2}(x)$ form a proper subset of the eigenvalues
of $T_{\rm RSOS}(x)$. While it is indeed true that the leading eigenvalue of
$T_{\rm RSOS}^{Z_2}(x)$ belongs to the spectrum of $T_{\rm RSOS}(x)$ (see
Table~\ref{tab1} for a numerical check), this inclusion is not necessarily
true for subdominant eigenvalues of $T_{\rm RSOS}^{Z_2}(x)$. 

Finally note that the leading eigenvalue of
$T_{\rm RSOS}^{Z_2}(x)$ coincides with a {\em subdominant} eigenvalue of
$T_{\rm RSOS}(x)$, as $Z_{\rm RSOS}^0(x)$ dominates $Z_{\rm RSOS}^{Z_2}(x)$.

In the case where $p/2$ is even, the difference $w_c^{\rm even}-w_c^{\rm odd}$ is 
the opposite between the twisted and untwisted models. Therefore, 
the conclusion is unchanged, except that the leading eigenvalue of $T_{\rm RSOS}^{{\rm even},Z_2}(x)$ 
coincides with a subdominant eigenvalue of $T_{\rm RSOS}^{\rm odd}(x)$, 
and the leading eigenvalue of $T_{\rm RSOS}^{{\rm odd},Z_2}(x)$
coincides with a subdominant eigenvalue of $T_{\rm RSOS}^{\rm even}(x)$.

\subsection{Twisted cluster model}

We want to extend the partition functions $Z^{Z_2}_{\rm spin}(x)$ and
$Z^{Z_3}_{\rm spin}(x)$, considered in section~\ref{sec2} by
twisting the spin representation for $Q=3$, to arbitrary values of 
$Q$. 
Within the cluster representation we introduce a 
horizontal seam. We define $Z_{\rm cluster}^{Q_0=1}(x)$   
by giving a weight $1$ to the non-trivial direct clusters with $i_2$ odd 
and to degenerate cycles with a direct cluster percolating,
while other direct
clusters continue to have the weight $Q$. We define too $Z_{\rm cluster}^{Q_0=0}(x)$ 
by giving a weight $0$ to the non-trivial direct clusters with $i_2$ coprime with $3$ 
and to degenerate cycles with a direct cluster percolating, while other  
direct clusters continue to have the weight $Q$.
$Z_{\rm cluster}^{Q_0=1}(x)$ 
and $Z_{\rm cluster}^{Q_0=0}(x)$ are extensions to  
arbitrary real values of $Q$ of, respectively, $Z^{Z_2}_{\rm spin}(x)$ and  
$Z^{Z_3}_{\rm spin}(x)$.  

We have the following duality relation for the $Q_0=1$ model:
\beq 
 Z_{\rm cluster}^{Q_0=1}(x) = x^E Z_{\rm cluster}^{Q_0=1}(x^{-1})
 \label{Q0=1}
\eeq
Indeed, for $Q_0=1$, the weight of a degenerate cycle is always 1,     
the cluster percolating being direct or dual.
That is the reason why this equality is true,
whereas it was false for the untwisted model because of the
degenerate Pasquier graphs. For $Q=3$, one retrieves Eq.~(\ref{duZ2}).

For the $Q_0=0$ model, there is a duality relation of the form:
\beq
Z_{\rm cluster}(x) - x^E Z_{\rm cluster}(x^{-1}) =
-(Q-1)\left(
Z_{\rm cluster}^{Q_0=0}(x) - x^E Z_{\rm cluster}^{Q_0=0}(x^{-1}) 
\right).
\eeq
Indeed, the two sides are non-zero only because of the degenerate
Pasquier graphs, so the relation follows by comparing weight of cycles
in the twisted and untwisted models. Combining this equation with
Eq.~(\ref{maindual}) enables us to relate the $Q_0=0$ twisted cluster
model to the RSOS model:
\beq
Z_{\rm RSOS}^{\rm even}(x) - x^E Z_{\rm RSOS}^{\rm odd}(x^{-1}) =
-\left(
Z_{\rm cluster}^{Q_0=0}(x) - x^E Z_{\rm cluster}^{Q_0=0}(x^{-1})
\right) \mbox{ for $p$ even}.
 \label{Q0=0}
\eeq
For $Q=3$, this reduces to Eq.~(\ref{duZ3}) as it should.

Note that these equations are correct because of the weight $Q_0$ chosen  
for degenerate cycles with a direct cluster percolating, so the partition
functions of other models, with the same value of $Q_0$ but differents weights 
for other configurations, would verify the same equations.

\subsection{The Ising case}

Let us now discuss in more detail the Ising case, $Q=2$ and $p=4$. In this
case, the relationship between the RSOS and spin pictures is actually trivial,
as the two transfer matrices are isomorphic. Consider for example $T_{\rm
RSOS}^{\rm even}(x)$. All direct clusters have height $h_i=2$, and dual
clusters have $h_i=1$ or $3$. The dual heights thus bijectively define Ising
spin variables $S_i=1$ or $2$ on the dual vertices.

To examine the weight of a lattice face, we decide to redistribute the
factor $Q^{L/2}$ in Eq.~(\ref{TM}) as a factor of $Q^{1/2}$ for each
$V_i$ operator, i.e., on faces which are like in Fig.~\ref{fig1}.
If $h_{2i}=h'_{2i}=1$ or $3$ the weight is then
$Q^{1/2} (x+S_1/S_2) = {\rm e}^K$, and if $h_{2i} \neq h'_{2i}$ the
weight is $Q^{1/2} S_1/S_2 = 1$. A similar reasoning holds on the faces
associated with an $H_i$ operator, this time with no extra factor of $Q^{1/2}$.
So these are exactly the weights needed to define an Ising model on the
dual vertices. Arguing in the same way in the odd RSOS sector, we conclude
that
\beq
 T_{\rm RSOS}^{\rm even}(x) = T_{\rm spin}(x), \qquad
 T_{\rm RSOS}^{\rm odd}(x) = x^{2L} T_{\rm spin}(x^{-1}),
 \label{IsingTM}
\eeq
cf.~Eq.~(\ref{Tspindual}).

Using again the explicit relation between heights and (dual) spins, the
$Z_2$ twist in the RSOS model is seen to be the standard $Z_2$ twist of
the Ising model (antiperiodic boundary conditions for the spins).
Since all local face weights are identical in the two models we have
as well
\beq
 T_{\rm RSOS}^{{\rm even},Z_2}(x) = T_{\rm spin}^{Z_2}(x), \qquad
 T_{\rm RSOS}^{{\rm odd},Z_2}(x) = x^{2L} T_{\rm spin}^{Z_2}(x^{-1}).
\eeq

\subsection{The $Q=3$ case}
\label{secQ3}

For $Q=3$, we have conjectured an additional relation, given by
Eq.~(\ref{comboZ2}), which we recall here:
\beq
 Z_{\rm RSOS}^{\rm even}(x) + Z_{\rm RSOS}^{{\rm even},Z_2}(x)=
 Z_{\rm spin}(x) + Z_{\rm spin}^{Z_2}(x)
 \label{Q3relation}
\eeq
This equation can be proved by considering the weights of all kinds of
Pasquier graphs that one might have. The contribution of non-degenerate
Pasquier graphs with clusters percolating vertically is $w = 
Q^{(l-n)/2}(Q^{n/2} + 1)$ on both sides. The contribution of non-degenerate
Pasquier graphs with clusters percolating horizontally is
$2 Q^{(l-n)/2}Q^{l/2}$. The contribution of degenerate Pasquier graphs is
$Q^{l/2} (Q+1)$ if the non-trivial cluster is direct, and
$2 Q^{l/2}$ if the non-trivial cluster is dual. Note that this
relation cannot be extended to other values of $p$, as we used the explicit
expressions for the eigenvalues of $G_5$ and that $p/2=Q$ and $(p-2)/2=2$.

\section{Discussion}
\label{sec4}

In this paper we have studied the subtle relationship between Potts
and RSOS model partition functions on a square-lattice torus. The
subtleties come from clusters of non-trivial topology, and in
particular from those that wind around {\em both} of the periodic
directions. Treating these effects by means of rigorous combinatorial
considerations on the associated Pasquier graphs has produced a number
of exact identities, valid on finite $L \times N$ tori and for any
value of the temperature variable $x = Q^{-1/2}({\rm e}^K-1)$. These
identities link partition functions in the RSOS and cluster
representations of the Temperley-Lieb algebra, in various parity
sectors and using various twisted versions of the periodic boundary
conditions.

Our main results are given in Eqs.~(\ref{anyparity}), (\ref{maindual}),
(\ref{diffZ2}), (\ref{Q0=1}), (\ref{Q0=0}) and (\ref{Q3relation}). At the
selfdual (critical) point $x=1$, some of these relations reduce to
tautologies, but taking derivatives of the general relations with respect to
$x$ before setting $x=1$ nevertheless produces non-trivial identities,
such as Eq.~(\ref{derividentity}).

Note that we have proved the identities on the level of partition functions,
but the fact that they are valid for any $N$ means that there are strong
implications for the eigenvalues of the transfer matrices. Let us write for a
given partition function $Z(x)=\sum_{i} \alpha_{i} \big( \Lambda_{i}(x)
\big)^N$, where (for a given width $L$) $\Lambda_{i}(x)$ are the eigenvalues
of the corresponding transfer matrix and $\alpha_{i}$ their multiplicities.
Consider now, as an example, Eq.~(\ref{maindual}) for $Q>1$ integer. Since it
is valid for any $N$, we have that the multiplicities satisfy
\beq
 (Q-1) \big( \alpha_{\rm RSOS}^{\rm even} - \alpha_{\rm RSOS}^{\rm odd} \big)
 = \alpha_{\rm spin} - \alpha_{\rm spin}^{\rm dual}
\eeq
for any eigenvalue. For instance, if
$\alpha_{\rm RSOS}^{\rm even} > \alpha_{\rm RSOS}^{\rm odd}$ we can conclude
that the corresponding eigenvalue also appears at least in $T_{\rm spin}(x)$,
and possibly also in $T_{\rm spin}^{\rm dual}(x)$ with a smaller multiplicity.
The possibility that for some eigenvalues
$\alpha_{\rm RSOS}^{\rm even} = \alpha_{\rm RSOS}^{\rm odd}$ explains why
only some but not {\em all} the eigenvalues of the transfer matrices
contributing to the individual terms in Eq.~(\ref{maindual}) coincide.
Similar considerations can of course be applied to our other identities.
For example, considering Eq.~(\ref{diffZ2}), one deduces that the 
eigenvalues
of the RSOS transfer matrix with a different multiplicity between the even  
and odd sectors are eigenvalues of the twisted RSOS transfer matrix.

The methods and results of this paper can straightforwardly be adapted to
other boundary conditions (e.g., with twists in both periodic directions) or
to other lattices (in which case the relations will typically link partition
functions on two different, mutually dual lattices).

It is of interest to point out the operator content of the twisted boundary
conditions that we have treated. We here consider only the case $x=1$ for which
the continuum limit of the RSOS model is \cite{Huse,PS} the unitary minimal
model ${\cal M}_{p,p-1}$. By standard conformal techniques, the ratio of
twisted and untwisted partition functions on a cylinder can be linked to two-point
correlation function of primary operators.

In the case of the $Z_2$ twist of the RSOS model (which is possible only for
even $p$) the relevant primary operator is $\phi_{p/2,p/2}$, the magnetic
operator of the Potts model, of conformal weight $h_{p/2,p/2} =
\frac{p^2-4}{16p(p-1)}$. To see this, first note that the argument given in
section~\ref{hieigtw} implies that the ratio of partition functions is
proportional to the probability that both endpoints of the $Z_2$ seam are
contained in the same (dual) cluster. As we are at the selfdual point ($x=1$)
we may as well refer to a direct cluster. Now this probability is proportional
to both the connected spin-spin correlation function (in the spin or cluster
formulations) and to the connected height-height correlation function (in the
RSOS formulation). The corresponding decay exponent is then the conformal
weight $h_{p/2,p/2}$ of the magnetisation operator.

For the special case of $p=6$ ($Q=3$) we have also discussed $Z_2$ and $Z_3$
type twists in the spin representation. The former is linked to the operator
$\phi_{2,2}$ of conformal weight $h_{2,2}=\frac{1}{40}$, and the latter is
linked to $\phi_{4,4}$ with $h_{4,4}=\frac{1}{8}$. The astute reader will
notice that both operators are actually not present in the 3-state Potts model
but belong to the larger Kac table of the minimal model ${\cal M}_{6,5}$. This
is consistent with the fact that the RSOS model with parameter $p$ is
precisely \cite{PS} a microscopic realisation of the minimal model ${\cal M}_{p,p-1}$.
In other words, the two types of twists generate operators which cannot be
realised by fusing local operators in the spin model, but are nevertheless
local operators in the RSOS model.

The operators $\phi_{2,2}$ and $\phi_{4,4}$ are most conveniently represented
as the two types of fundamental disorder operators \cite{ZamFat} in the $Z_3$
symmetric parafermionic theory (the coset $su(2)_3/u(1)$) which is an extended
CFT realisation of ${\cal M}_{6,5}$. More precisely, in the notation of
Eq.~(3.38) in Ref.~\cite{ZamFat} we have $h_{4,4}=\Delta_{(0)}=\frac{1}{8}$
and $h_{2,2}=\Delta_{(1)}=\frac{1}{40}$.

Finally note that $T_{\rm RSOS}^{Z_2}$ contains levels which are not
present in $T_{\rm RSOS}$, cf.~Table~\ref{tab1}. Thus, at $x=1$ the
corresponding operator content is different from that of ${\cal M}_{p,p-1}$.
In particular, correlation functions must be defined on a two-sheet Riemann
sphere, and we expect half-integer gaps in the spectrum. This expectation is
indeed brought out by our numerical investigations: for $p=6$ the second
scaling level in the twisted sector is a descendant of $\phi_{2,2}$ at
level $1/2$.
These issues will be discussed further elsewhere.

\end{document}